A Modest View of Bell's Theorem

Steve Boughn, Princeton University and Haverford College

Talk given at the 2016 *Princeton-TAMU Symposium on Quantum Noise Effects in Thermodynamics, Biology and Information*

In the 80 years since the seminal 1935 Einstein-Podolsky-Rosen paper (EPR), physicists and philosophers have mused about what Einstein referred to as "spooky action at a distance". Bell's 1964 analysis of EPR-type experiments has been cited more than 10,000 times and I suspect that most of the citations have occurred in the last decade. Many of these describe experimental work on realizing the quantum mechanical predictions related to entangled states and, as an experimentalist, I have great admiration for much of this work. On the other hand many, if not most, of the remainder are theoretical and philosophical papers trying to come to grips with what is often referred to as the nonlocality of quantum mechanics. Bell's conclusion was that any hidden variable theory designed to reproduce the predictions of quantum mechanics must necessarily be nonlocal and allow superluminal interactions. In his words

> In a theory in which parameters are added to quantum mechanics to determine the results of individual measurements, without changing the statistical predictions, there must be a mechanism whereby the setting of one measurement device can influence the reading of another instrument, however remote. Moreover, the signal involved must propagate instantaneously, so that such a theory could not be Lorentz invariant.

This doesn't immediately imply that nonlocality is a characteristic feature of quantum mechanics let alone a fundamental property of nature; however, many physicists and philosophers of science do harbor this belief. In fact, Bell (1975) later described the violation of his inequalities as pointing to the "gross non-locality of nature." Statements like "If one assumes the world to be real, then Bell-experiments have proven that it is nonlocal" (Wiseman 2006) and "Bell's theorem asserts that if certain predictions of quantum theory are correct then our world is non-local…experiments thus establish that our world is non-local" (Goldstein et al. 2011) abound in scientific literature, and especially in popular literature. Experts in the field often use the term "nonlocality" to designate particular non-classical aspects of quantum entanglement and do not confuse the term with superluminal interactions. However, many physicists seem to take the term more literally.

For me the term, *non-locality*, is so fraught with misinterpretation that I feel we'd all be better off if it were removed from discourse on quantum mechanics. I confess that I'm neither a theoretical physicist nor a philosopher but rather an experimental physicist and observational astronomer and it's from this vantage point that I'll try to convince you that quantum mechanics does not require spooky action at a distance of any kind and will then tell you to what I attribute the magic of quantum mechanical entanglement.

The conclusion of EPR was not that quantum mechanics is non-local but rather that quantum mechanics is not a complete description of reality. Einstein based his argument on what he termed the "separation principle", according to which the real state of affairs in one part of space cannot be affected instantaneously or super-luminally by events in a distant part of space. In a 1935 letter to Schrödinger, Einstein explained (Howard 2007)

> After the collision, the real state of (*AB*) consists precisely of the real state *A* and the real state of *B*, which two states have nothing to do with one another. *The real state of B thus cannot depend upon the kind of measurement I carry out on A* [separation hypothesis]. But then for the same state of *B* there are two (in general arbitrarily many) equally justified $\Psi_B$, which contradicts the hypothesis of a one-to-one or complete description of the real states.

In some ways, I completely agree with the separation principle; however, from my experimentalist perspective, I would interpret it as "If systems A and B are spatially separated, then a measurement of system A can, in no way, have any effect on any possible measurement of system B," a principle with which, I trust, few would disagree. As an illustration, consider Bell's original gedanken experiment (actually due to Bohm and Aharonov - 1957): the emission of two oppositely moving spin ½ particles in a singlet state. Their combined wave function is given by

$$\Psi(1,2) = \frac{1}{\sqrt{2}}\{|1,\uparrow\rangle|2,\downarrow\rangle - |1,\downarrow\rangle|2,\uparrow\rangle\}_z$$

where ↑ and ↓ indicate the *z* components of the spins of particles 1 and 2. Now suppose that the spin of particle 1 is measured with a Stern-Gerlach apparatus oriented in the $\hat{z}$ direction and is determined to be ↑. The usual statement is that such a measurement instantaneously collapses the wave function of particle 2 such that $\Psi(2) = |2,\downarrow\rangle_z$. On the other hand, the original wave function can also be expressed as

$$\Psi(1,2) = \frac{1}{\sqrt{2}}\{|1,\uparrow\rangle|2,\downarrow\rangle - |1,\downarrow\rangle|2,\uparrow\rangle\}_x.$$

Then, if the spin of particle 1 is measured with a Stern-Gerlach apparatus oriented in the $\hat{x}$ direction and is determined to be ↑, the wave function of particle 2 collapses to $\Psi(2) = |2,\downarrow\rangle_x$. This violates Einstein's separation principle that the measurement of particle 1 can have no effect on the state of particle 2. On the other hand, this scenario does not violate my version of the separation principle. If one measures particle 1 to be ↑ in any direction, we know the measurement of particle 2 has to be ↓. This perfect (anti)correlation is built into the two particle system because they are in a singlet state. The problems becomes a bit stickier if one measures the spin of particle 1 in an arbitrary direction $\hat{n}$ where $\hat{n}\cdot\hat{z} = \cos\theta$. It is convenient to express $|1,\uparrow\rangle_z$ and $|1,\downarrow\rangle_z$ in an $\hat{n}$ basis, i.e

$$|1,\uparrow\rangle_z = \cos\tfrac{\theta}{2}|1,\uparrow\rangle_n + \sin\tfrac{\theta}{2}|1,\downarrow\rangle_n$$

and

$$|1,\downarrow\rangle_z = -\sin\tfrac{\theta}{2}|1,\uparrow\rangle_n + \cos\tfrac{\theta}{2}|1,\downarrow\rangle_n.$$

Then it is straightforward to show that the correlation of the measured components of the spins of the two particles in the $\hat{z}$ and $\hat{n}$ directions is given by $-\cos\theta$ (with the spins in units of $\hbar/2$). That is, the correlations between the measurements of the two distant systems are changed, which may seem to be a problem even for my separation principle.

How is it that the original states of the two particles know about the perfect anti-correlation, this new correlation and, in fact, about every correlation of all conceivable measurements made on the two particles? Well, that's the magic of quantum mechanical entanglement but it does not violate my version of the separation principle. Imagine that the measurement of particle 2 occurs well before that of particle 1. (For example, particle 2 is much closer to the initial interaction region than is particle 1.) Now suppose that the result of that measurement is sent via a light signal to the observer of particle 1 and is recorded in a lab notebook by an assistant well before the choice of the direction $\hat{n}$ is made. Then it is absolute clear that that choice of measurement of 1 could have no possible effect on the measurement of 2, superluminal or otherwise, as is consistent with my version of the separation principle. After all, the measurement is "written in stone". The correlations of many such measurements will be those predicted by quantum mechanics regardless of the choice of measurements. All those correlations are built into the entangled wave function of the two particles. As for the measurements of particle 1 and 2 themselves, they have absolutely no effect on each other. There is no spooky action at a distance.

Okay, so how do I explain the magic of the correlations of entangled quantum mechanical states? First, a quick review of Bell's theorem. Let $P(\hat{a}, \hat{b})$ denote the correlation of spin measurements of particles 1 and 2 along the directions $\hat{a}$ and $\hat{b}$ respectively (Bell's notation). Then the inequality derived by Bell (roughly a one page proof), $1 + P(\hat{b}, \hat{c}) \geq |P(\hat{a}, \hat{b}) - P(\hat{a}, \hat{c})|$, is the condition that must be satisfied by any classical, local, hidden variable description of the hypothetical singlet spin system. Suppose that $\hat{b}$ is in the $\hat{z}$ direction while $\hat{a}$ and $\hat{c}$ are in directions that are $\pm\theta$ from the $\hat{z}$ direction. Then for the quantum mechanical correlation analysis above, Bell's inequality takes the form $1 - \cos\theta \geq |\cos 2\theta - \cos\theta|$, which is clearly violated for $0 < \theta < \frac{\pi}{2}$ (the domain of applicability of the inequality). The implication is that quantum mechanics is inconsistent with any classical, local, hidden variable theory.

The violation of the Bell inequality arises from the quantum mechanical prediction of the correlations of measurements of the spins, i.e., $P(\hat{z}, \hat{n}) = -\cos\theta$, but this correlation arises directly from the single particle wave function. Suppose a spin ½ particle is in the $|\uparrow\rangle_z$ state. From the above relation, this can also be written in the $\hat{n}$ basis as $|\uparrow\rangle_z = \cos\frac{\theta}{2}|\uparrow\rangle_n + \sin\frac{\theta}{2}|\downarrow\rangle_n$. Now consider the correlation between a measurement of the particle's spin in the $\hat{z}$ direction and a hypothetical measurement of the spin of the same particle in the $\hat{n}$ direction. Then

$$P(\hat{z}, \hat{n}) = \langle\uparrow|_z \sigma_z \sigma_n \left(\cos\frac{\theta}{2}|\uparrow\rangle_n + \sin\frac{\theta}{2}|\downarrow\rangle_n\right) = \cos\theta.$$

This is the same correlation as for Bell's gedanken experiment except for the minus sign, which is due to the singlet state of Bell's two particles. One can easily make them exactly the same simply by requiring either of the measurements to change the sign of the output. So it seems that this hypothetical measurement also violates Bell's inequality even though there is no question of non-locality or superluminal interactions. After all, there's only a single particle. Of course, the measurement of the spin in the $\hat{n}$ direction is only hypothetical because one can't simultaneously measure components corresponding to non-commuting operators. Philosophers refer to the results of such a measurement as *counterfactually definite*. Bell's theorem clearly makes use of counterfactual definiteness; his inequality involves the correlations of the spins of the two particles in each of two different directions that correspond to non-commuting spin components. Even so, I realize that one might object to such a counterfactual gedanken experiment. However, the purpose of this hypothetical experiment is to demonstrate that the resulting correlation has little to do with entanglement, non-locality, and superluminal signals and everything to do with superposition and the nature of quantum states, whether entangled or not. In fact, one can render the counterfactual measurement to be an actual measurement by cloning the original single particle state and performing an actual measurement on the cloned state. In effect, this is precisely what Bell's singlet entangled state accomplishes.

Finally, consider the same single particle state but this time correlate two actual measurements of the spin in the $\hat{z}$ and $\hat{n}$ directions. The experimental arrangement is a Stern-Gerlach apparatus aligned in the $\hat{z}$ direction with another Stern-Gerlach apparatus located in the upper arm of the former but aligned in the $\hat{n}$ direction (see Fig. 1). Electron counters, $D_1$ and $D_2$, are placed at the two outputs of the $2^{nd}$ apparatus. If there is a detection in either of these two outputs, then we know that the state of the electron incident on the 2$^{nd}$ apparatus is $|\uparrow\rangle_z$ and so can consider this to be a measurement of the $\hat{z}$ component of the spin of the incident electron. On the other hand, $|\uparrow\rangle_z$ can be represented as $|\uparrow\rangle_z = \cos\frac{\theta}{2}|\uparrow\rangle_n + \sin\frac{\theta}{2}|\downarrow\rangle_n$ and so the specific counter that registers the detection can be considered a measurement of the $\hat{n}$ component of the spin. Then the correlation of the spin of the incident electron, +1, with the spin detected by the second apparatus is again given by $P(\hat{z},\hat{n}) = \cos\theta$. Thus one can argue that the correlation of these two measured values of the spin again violates the Bell inequality. In this case the final step of the two measurements occur simultaneously and so non-locality and superluminal wave function collapse are irrelevant.

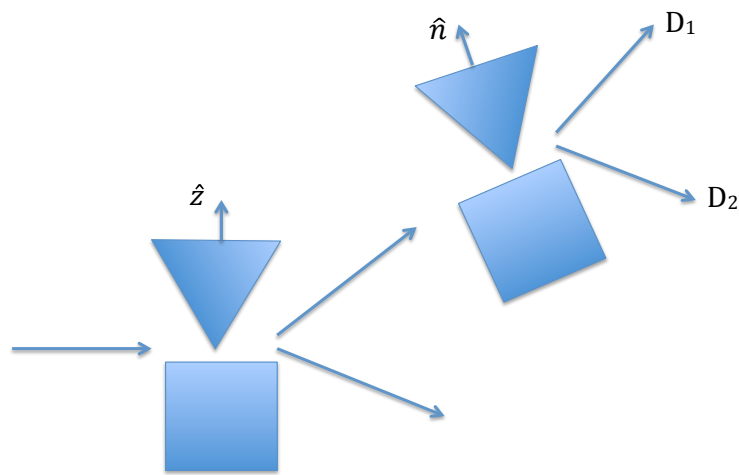

Figure 1: Double Stern-Gerlach Apparatus

The phenomenon of entanglement is not restricted to quantum mechanics. Two classical particles that interact with each other before moving off in different directions are also entangled. To the extent that the interaction can be completely characterized, one can predict the correlations of all possible measurements made on the two particles. On the other hand, there is certainly no doubt that quantum entanglement is a much richer phenomenon than its classical counterpart, cf. quantum information and quantum computing. However, we all know that quantum mechanics is, in general, much richer than classical mechanics. Quantum theory is capable of describing atoms and their interactions, the properties of solids, liquids, and gases, and is applicable not only to physical systems but to chemistry and, by inference, biology as well, whereas classical mechanics has had little to offer for such systems. Quantum entanglement, as Mermin declared (1985), "…is as close to magic as any physical phenomenon… and magic should be enjoyed." What I have endeavored to demonstrate is that Bell's theorem and what it reveals about entangled states have nothing to do with quantum non-locality and superluminal propagation. Rather, they follow directly from the single particle quantum mechanical superposition, $|\uparrow\rangle_z = \cos\frac{\theta}{2}|\uparrow\rangle_n + \sin\frac{\theta}{2}|\downarrow\rangle_n$. The simple examples above are intended to demonstrate that the correlations of entangled states can be understood in terms of the standard quantum correlations of any system whether entangled or not.

The conclusion of the EPR paper was neither that quantum mechanics yields incorrect predictions nor that quantum mechanics is non-local. In fact, it was the apparent presence of super-luminal effects that led Einstein to proclaim that quantum mechanics does not provide a complete description of reality. I agree that quantum mechanics is not a complete description of reality, albeit for a different reason. The EPR and Bell papers both invoke the notion of wave function collapse that accompanies the act of measurement. The problem is that wave function collapse is not predicted by quantum mechanics. It is an event that happens outside the mathematical formalism of the theory. In fact, formal quantum theory says absolutely nothing at all about measurements and how they should be performed or even how to interpret the results of measurements. The statistical distributions of quantum mechanical predictions follow from rules for how to interpret the *constructs* of quantum theory in terms of the results of experiments. The measurements themselves are not described by quantum theory but rather by the operational prescriptions of the experimentalists and technicians who perform them. It is in this sense that quantum mechanics is incomplete. Einstein arrived at his conclusion from his separation principle, as described above. I would characterize his mistake (and I say this with trepidation!) as choosing to characterize "reality" by a mathematical model, a path that can easily lead one astray.